\documentclass[sn-mathphys-num]{sn-jnl}

\usepackage{graphicx}%
\usepackage{multirow}%
\usepackage{amsmath,amssymb,amsfonts}%
\usepackage{amsthm}%
\usepackage{mathrsfs}%
\usepackage[title]{appendix}%
\usepackage{xcolor}%
\usepackage{textcomp}%
\usepackage{manyfoot}%
\usepackage{booktabs}%
\usepackage{algorithm}%
\usepackage{algorithmicx}
\usepackage{algpseudocode}%
\usepackage{listings}%
\usepackage{makecell}%
\usepackage{indentfirst}%

\theoremstyle{thmstyleone}%
\theoremstyle{thmstyletwo}%

\theoremstyle{thmstylethree}%

\begin{document}

\title[Article Title]{Design and Implementation of TAO DAQ System}


\author*[1,2]{\fnm{Shuihan} \sur{Zhang}}\email{zhangshuihan@ihep.ac.cn}
\author[1,2]{\fnm{Chao} \sur{Chen}}
\author*[2]{\fnm{Xiaolu} \sur{Ji}}\email{jixl@ihep.ac.cn}
\author[2]{\fnm{Fei} \sur{Li}}
\author[1,2]{\fnm{Yu} \sur{Peng}}
\author[3]{\fnm{Fabrizio} \sur{Petrucci}}
\author[1,2]{\fnm{Yinhui} \sur{Wu}}
\author[2]{\fnm{Zezhong} \sur{Yu}}
\author[2]{\fnm{Tingxuan} \sur{Zeng}}
\author[1,2]{\fnm{Kejun} \sur{Zhu}}

\affil[1]{ \orgname{University of Chinese Academy of Sciences}, \orgaddress{ \city{Beijing}, \postcode{100049}, \country{China}}}

\affil[2]{ \orgdiv{State Key Laboratory of Particle Detection and Electronics}, \orgname{Institute of High Energy Physics, Chinese Academy of Sciences}, \orgaddress{\city{Beijing}, \postcode{100049}, \country{China}}}
\affil[3]{ \orgdiv{Istituto Nazionale di Fisica Nucleare Sezione di Roma Tre}, \orgname{Roma}, \country{Italy}}


\abstract{The abstract serves both as a general introduction to the topic and as a brief, non-technical summary of the main results and their implications. Authors are advised to check the author instructions for the journal they are submitting to for word limits and if structural elements like subheadings, citations, or equations are permitted.}


\abstract{\textbf{Purpose:} The Taishan Antineutrino Observatory (TAO) is a satellite experiment of the Jiangmen Underground Neutrino Observatory (JUNO), also known as JUNO-TAO. Located close to one of the reactors of the Taishan Nuclear Power Plant, TAO will measure the antineutrino energy spectrum precisely as a reference spectrum for JUNO. The data acquisition (DAQ) system is designed to acquire data from the TAO readout electronics and process it with software trigger and data compression algorithms. The data storage bandwidth is limited by the onsite network to be less than 100 Mb/s.

\textbf{Methods:} The system is designed based on a distributed architecture, with fully decoupled modules to facilitate customized design and implementation. It is divided into two main components: the data flow system and the online software. The online software serves as the foundation, providing the electronics configuration, the process management, the run control, and the information sharing. The data flow system facilitates continuous data acquisition from various electronic boards or trigger systems, assembles and processes raw data, and ultimately stores it on the disk.

\textbf{Results:} The core functionality of the system has been designed and developed. The usability of the data flow system interface and the software trigger results have been verified during the pre-installation testing phase.

\textbf{Conclusion:} The DAQ system has been deployed for the TAO experiment. It has also successfully been applied to the integration test of the detector and electronics prototypes.}

\keywords{Data acquisition system, Data flow, Online, TAO}



\maketitle

\section{Introduction}\label{sec1}

The Taishan Antineutrino Observatory (TAO or JUNO-TAO) is a satellite experiment of the Jiangmen Underground Neutrino Observatory (JUNO). It will be deployed near a reactor core of the Taishan Nuclear Power Plant in Guangdong, China. The primary physics goal of TAO is to measure the reactor antineutrino energy spectrum with an unprecedented energy resolution better than 2\% at 1MeV \cite{abusleme2020tao} to provide a high-precision reference spectrum for JUNO, and provide a benchmark for the nuclear database. To achieve its goals, the experiment will realize an optical coverage of the 2.8 tons of Gd-loaded liquid scintillator (LS) close to 95\% with novel silicon photomultipliers (SiPMs), with a photon detection efficiency (PDE) above 50\% \cite{steiger2022tao}.

\begin{figure}[b]
    \centering
    \includegraphics[width=1\linewidth]{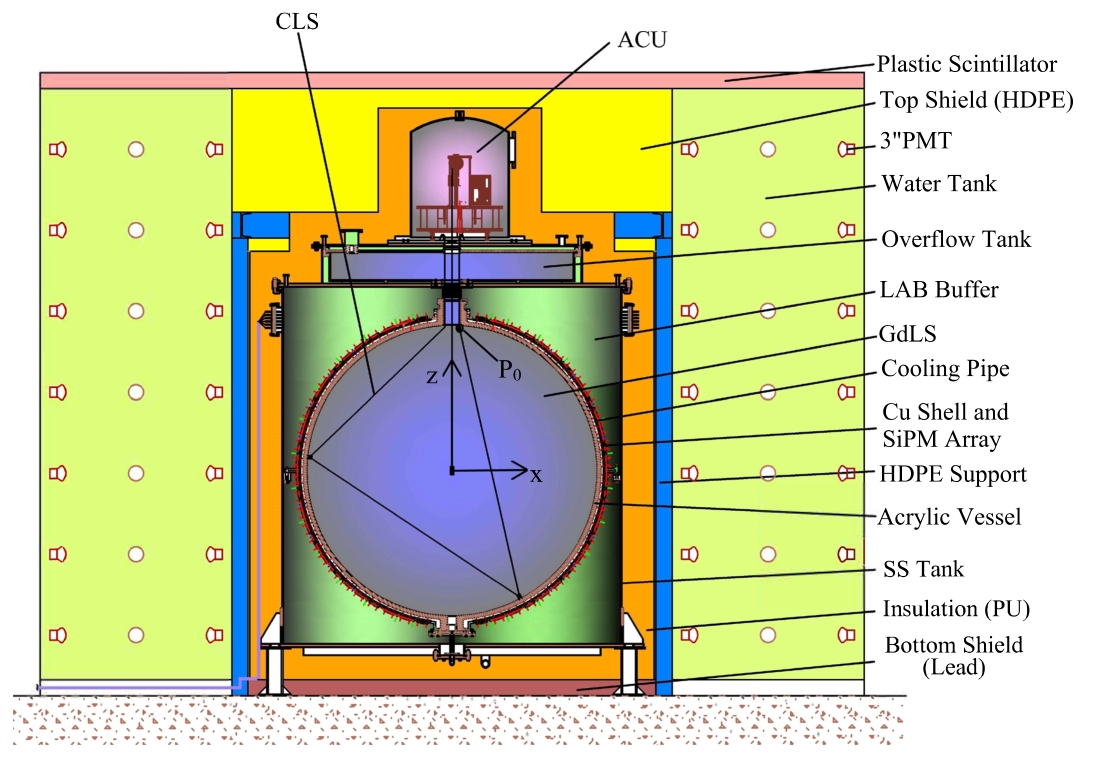}
    \caption{The conceptual design of the TAO detector \cite{xu2022calibration}}
    \label{fig1_detector}
\end{figure}

The conceptual design of the TAO detector is shown in Figure~\ref{fig1_detector}. TAO consists of the central detector, the calibration system, the outer shielding, and the veto system. In terms of data acquisition, TAO comprises three independent systems: a central detector (CD), a water tank (WT), and a top veto tracker (TVT) \cite{steiger2022tao}.

\begin{figure}[h]
    \centering
    \includegraphics[width=1\linewidth]{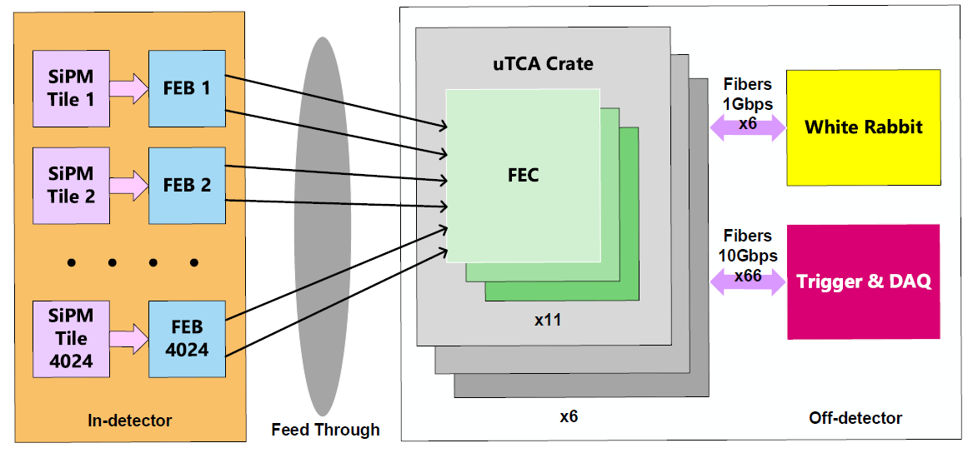}
    \caption{The architecture of the CD readout electronics system \cite{xieyuguang}}
    \label{fig_cdreadout}
\end{figure}

The TAO electronics system has been specifically designed to meet the unique requirements of each detector. For CD, the front-end readout is based on a multi-level distributed FPGA architecture. The overall architecture of the readout electronics system \cite{xieyuguang} is shown in Figure \ref{fig_cdreadout}. The analog signals generated by the SiPM are amplified and shaped by the Front End Boards (FEBs) and then sent to the Front End Controllers (FECs). FECs pre-process and format the data, and then transmit the data to the trigger and data acquisition system (TDAQ). The TVT utilizes a multi-layer strip plastic scintillator array. The electronics system is responsible for processing signals from the SiPMs, which includes signal amplification, analog signal digitization, time/charge measurement, and data transmission. Subsequently, the TVT trigger system performs channel hits coincidence and trigger selection in both channels of a single plastic scintillator strip, while the DAQ system performs coincidence trigger selection between multi-layer scintillators. The WT electronics system is based on the design of the JUNO 3-inch photomultiplier (PMT) \cite{adam2015juno}, with the Global Control Unit (GCU) \cite{junolpmt1, CERRONE2023168322}, serving as its central component. The GCU is connected to the DAQ system via Ethernet.

The data acquisition (DAQ) system is responsible for collecting data fragments generated by the electronics and trigger systems in the experiment, assembling and processing them into events, and finally storing them on the disk. Meanwhile, the running status and log information need to be monitored and recorded. Therefore, the DAQ system will directly affect the reliability and performance of the system, and it requires high processing efficiency without additional dead time.
The data readout interface and data type are different for each detector system as shown in Table~\ref{tab1_feactures}.

-	For CD, the triggered data from the Central Unit (CU) will be collected by the DAQ system, and the data bandwidth input to the DAQ system is approximately 800 Mb/s \cite{liu2023nhit,abusleme2020tao, jixl}. Therefore, to ensure redundancy and stability, SiTCP \cite{sitcp} protocol and 10 Gigabit Ethernet (GbE) have been chosen as the readout interface. This link supports the acquisition of time and charge (T/Q) for each CD channel.

-	For WT, the same electronics system as JUNO 3-inch PMT is used. The DAQ receives data from GCUs, and the TCP-based readout method and the IPbus-based \cite{2016ipbus} configuration method have been used as the data transmission protocol. In total, three GCUs are employed. Only T/Q hits information is supported.

-	For TVT, the DAQ collects data from the TVT Gather Unit (TGU), using the SiTCP protocol and 1 Gigabit Ethernet as the data interface. Only T/Q information is included.

\begin{table}[h]
\caption{DAQ Readout interface features for TAO experiment}\label{tab1_feactures}%
\begin{tabular}{@{}llll@{}}
\toprule
Detector & Interface  & Connection boards & Data Type\\
\midrule
CD    & 10GbE+SiTCP  & 1 CU  & T/Q hits  \\
WT    & GbE+TCP/IPbus   & 128 PMTs/GCU  & T/Q hits  \\
TVT    & GbE+SiTCP   & 2 TGUs  & T/Q hits  \\
\botrule
\end{tabular}
\end{table}

This paper introduces the design and implementation of the TAO DAQ system including software architecture and hardware deployment. In addition, an overview of the progress currently being made in the R\&D for the TAO DAQ system will be presented.

\section{System requirements}\label{sec2}

\subsection{Readout throughput requirements}\label{subsec2}

Based on the design specification of the TAO experiment \cite{abusleme2020tao, xieyuguang}, the trigger system of the CD effectively reduces SiPM dark noise and radioactive backgrounds, resulting in a data input rate to the DAQ system of less than 1 Gb/s. The WT, consisting of 300 PMTs and utilizing the same electronics design as the JUNO 3-inch PMT, has a total data rate of approximately 105 Mb/s. The TVT, which comprises 160 plastic scintillator (PS) strips with 320 FEC channels, requires a data bandwidth from TGU to the DAQ system of around 40 Mb/s.

As a result, the DAQ system needs to be designed to handle a data readout bandwidth in the Gbps range and should support data transfer with negligible dead time.
\subsection{Data processing requirements}\label{subsec2}

As previously stated, there are multiple readout links between the DAQ and the electronics or trigger system. Therefore, the DAQ system must be capable of assembling data fragments from various readout links based on timestamp or trigger numbers, merging events from different detectors, and sorting them by timestamp. Additionally, due to the network data transmission bandwidth limitations, the system’s output bandwidth should be kept below 100 Mb/s. Real-time data compression and software triggering are essential to reduce the data rate. Furthermore, a careful study of the data processing algorithm is necessary to ensure that the system's performance meets the requirements of the TAO experiment within the limited online CPU resources.

\subsection{Other common requirements}\label{subsec2}

For high-energy physics experiments, the main task of the DAQ system is to record all kinds of data from each detector on disk. The DAQ system must offer essential functions such as electronics configuration, data readout, data processing, and data storage.

In addition, the system should provide common functions such as software configuration, run control, information sharing, and data quality monitoring. It is also important for the system to support long-term stable operation and be resilient to network fluctuations.

\section{Software architecture design}\label{sec3}

Considering the similarities between the JUNO and TAO experiments, as well as joint maintenance and analysis advantages, the TAO DAQ is designed and developed based on the framework of the JUNO DAQ \cite{adam2015juno, jixl_juno}, with customizations to meet its specific needs. This framework employs a plugin-based modular design, extracting common modules and components that can be easily reused by other experiments. The framework itself only provides overall business processing and scheduling, while each specific function is completed in plugin form. Users can customize the design and development of readout plugins and processing plugins according to their needs. In addition, the framework has been designed to support high availability. However, considering the convenience of system deployment and the requirements for easy maintenance of the TAO experiment, the feature is currently deactivated. If there is a need for high availability in the future, it can be easily enabled at any time.

\begin{figure}
    \centering
    \includegraphics[width=1\linewidth]{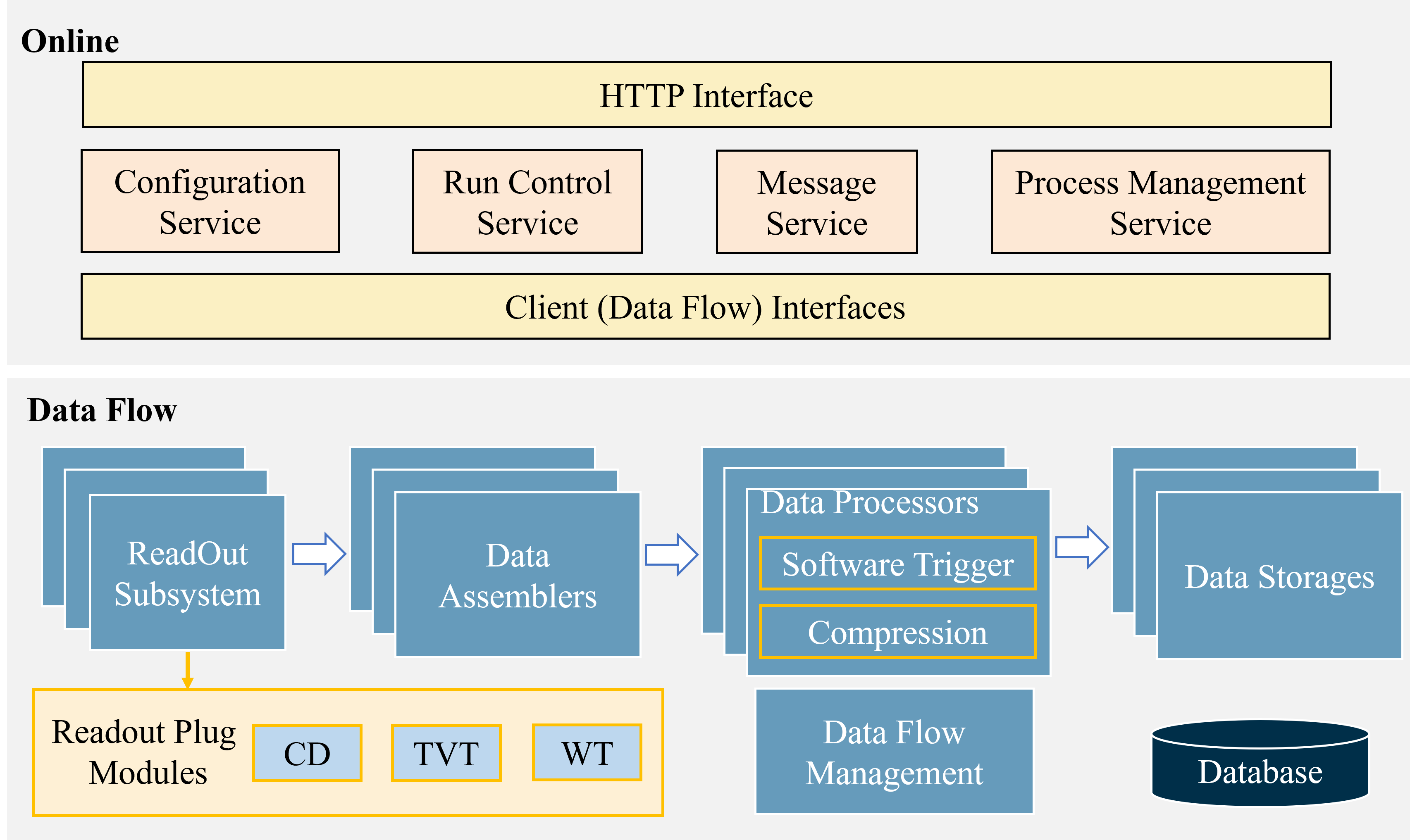}
    \caption{The architecture of TAO DAQ system}
    \label{fig2_artitecture}
\end{figure}

The TAO DAQ software design structure, as illustrated in Figure~\ref{fig2_artitecture}, is based on a distributed system architecture with fully decoupled modules. It comprises two main parts: the data flow system and the online software, which collaborate to achieve comprehensive functionality.

\begin{itemize}
\item \textbf{The data flow part} \cite{zeng2018juno, chenchao} serves as the core of the DAQ system, primarily responsible for retrieving data from the front-end electronics or trigger systems, customizing data processing algorithms based on the requirements of the TAO experiment, and packaging data into events for storage.

\item \textbf{The online software part} \cite{wuyinhui} is the structure foundation of the DAQ system, responsible for managing, controlling, and reporting messages for the data flow system. This component is utilized for configuration, run control, online monitoring, real-time display, and information sharing, among other functions.
\end{itemize}

The independent development and operation of the two components can effectively reduce inter-module coupling, minimize mutual interference, and enhance the overall system's flexibility and stability.

\section{System implementation and status}\label{sec4}

\subsection{Data flow}\label{subsec2}

As previously mentioned, the TAO DAQ system requires data acquisition from different detectors and readout electronics, with differences in data format and data processing algorithms. Therefore, it is necessary to customize the implementation of the data flow system.

\subsubsection{Data readout and configuration}\label{subsubsec2}
To simultaneously acquire data from three distinct detectors, the DAQ system provides data readout and configuration interfaces based on the SiTCP protocol and the IPbus protocol. It also supports parallel data readout and checks for different data formats. For abnormal data formats, we discard the anomalous data packets to facilitate the subsequent data assembly and processing. To facilitate development and maintenance, independent data readout and configuration modules for the three detectors have been implemented. Users can configure the data readout method through online software, after which the data flow system can access the configuration to execute the corresponding data readout.

\subsubsection{Data processing}\label{subsubsec2}

To achieve reliable data transmission and permanent storage, the DAQ system must minimize the data storage volume during the data processing phase. To address this, the TAO DAQ has implemented two methods: software trigger and data compression, to alleviate the burden on final data transmission.

\begin{itemize}
\item \textbf{Software trigger} 

The software trigger algorithm is commonly employed to identify and select events of interest from massive and complex data, thereby reducing the data rate that needs to be stored. TAO utilizes software triggers for WT and TVT, offering greater flexibility to modify algorithms. Furthermore, the framework of the data flow system provides interfaces for data processing algorithms, allowing for customization and simplifying maintenance upgrades. Currently, it primarily provides the following two software trigger algorithms: 

\textbf{- Multiplicity (nHit) trigger} The WT will utilize it to decrease the data bandwidth. Each PMT will be triggered separately with configurable thresholds. If the number of hit channels within a trigger window exceeds the specified threshold, it is considered a trigger and packaged into a complete event.

\textbf{- Layer coincidence trigger} The TVT employs it for software trigger. In this process, data from the front-end trigger board is acquired and divided into four layers based on specific mapping tables. A trigger is activated when at least two out of the four layers are hit, and all hits within the designated readout window are packaged into a single physical event.

Currently, two available algorithms have been provided: the “sorted hit” algorithm \cite{lu2020lhaaso} and the “tabled hit” algorithm \cite{pengyu_softwaretrigger}. Both algorithms offer enhanced flexibility through configurable thresholds, trigger windows, and readout windows. To validate the correctness of the software trigger algorithm, we performed a cross-check of the results of two algorithms using data from the Large High Altitude Air Shower Observatory (LHAASO) experiment \cite{lhaaso_nature}. The LHAASO experiment has no global hardware trigger system and data processing is mainly realized based on software trigger \cite{lhaaso}. The “sorted hit” algorithm has been applied in the LHAASO experiment and has been validated \cite{lu2020lhaaso}. Therefore, we input the same raw data into two algorithms and compared the results of the triggered events. Through comparison, we found that the output results of the two algorithms were completely consistent, verifying the feasibility of the algorithm.

Moreover, events generated from TVT simulations were utilized as the data source input for the data flow system to validate the correctness of the layer coincidence trigger algorithm. During the verification process, our primary focus was on checking for false triggers and missed triggers. A false trigger means that an event with the number of hit channels below the threshold in the trigger window is output, while a missed trigger means that an event with the number of hit channels exceeding the threshold in the trigger window is not output. We compared the results of the software trigger algorithm with the simulated results, as shown in Figure~\ref{fig_triggerver1}. We did not identify any false triggers or missed triggers, thereby confirming the accuracy of the software trigger algorithm.


\begin{figure}
\begin{minipage}{0.5\textwidth}
  \centering
  \includegraphics[width=1\linewidth]{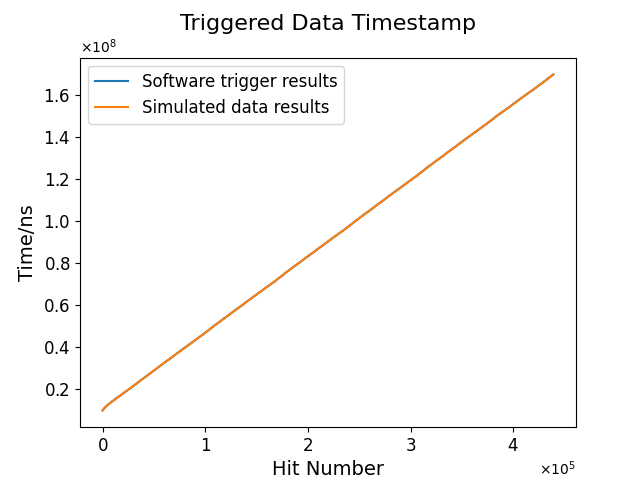}

\end{minipage}%
\begin{minipage}{0.5\textwidth}
  \centering
  \includegraphics[width=1\linewidth]{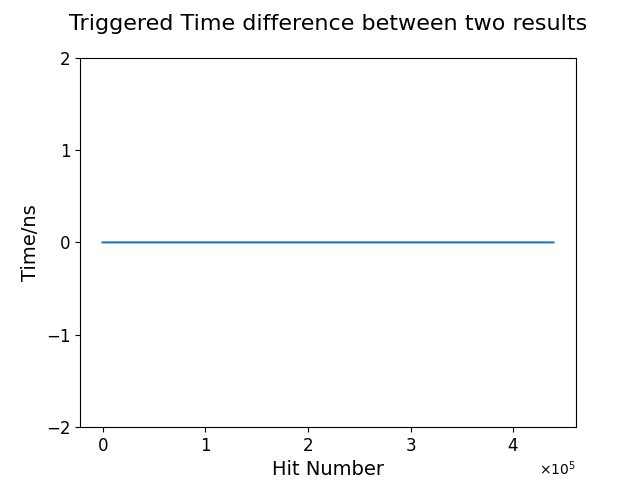}
\end{minipage}
\caption{The comparison between the software trigger algorithm and the simulated results}
\label{fig_triggerver1}
\end{figure}

\item \textbf{Data compression}

Using software for data compression provides increased flexibility and scalability without disrupting the existing architecture. Two data compression methods have been provided based on the specific data format and requirements of the TAO experiment.

\textbf{- Lossless compression} It is a data compression technique that reduces data size without any loss of information, allowing for complete restoration of the raw data. The CD and WT will further reduce the data rate using this method while ensuring data integrity. The data flow system has provided an algorithm interface to facilitate the integration of various compression algorithms. At the same time, modular design contributes to the updating and maintenance of algorithms.

\textbf{- Lossy compression} For CD, which handles a large data volume input to the DAQ system, it will additionally utilize online lossy data compression algorithms. Redundant information within data packets will be merged and reconstructed. These data compression techniques are expected to further reduce data transmission bandwidth and storage space while ensuring the security and reliability of disk data.
\end{itemize}
\subsubsection{Data storage}\label{subsubsec2}

The data storage (DS) module sorts the data fragments from three detectors by timestamp, merging and storing them in a disk array. To facilitate offline data analysis, file index information is stored in a database for easy retrieval. Furthermore, TAO will adopt the same data file format and data transmission system as JUNO. Simultaneously, the data will be continuously transmitted to a permanent storage platform to ensure secure and reliable long-term data storage.

TAO DAQ has completed the core development of its full-chain data flow system, providing support for different data readout and configuration interfaces: SiTCP-based protocol and IPbus-based protocol. Additionally, the data processing module also offers corresponding interfaces, supporting customized software trigger and compression algorithms.

\subsection{Online software}\label{subsec2}

The online software for TAO has integrated the existing online architecture from JUNO, including configuration, run control, and message services. Specifically, due to the relatively small scale of the TAO experiment compared to the JUNO experiment, and considering the need for convenient system deployment and easy maintenance, we have opted to use automated scripts for process management instead of utilizing Kubernetes \cite{k8s}. In addition, customized online monitoring has adapted to the specific characteristics of the TAO experiment, offering enhanced support for data collection and processing.

The monitoring indicators are categorized into hardware and software dimensions. Hardware monitoring involves tracking the system status of servers, switches, and disks to promptly detect any potential anomalies. In terms of software, it's crucial to monitor the data flow system, record the running status, CPU and memory usage, and provide visualization tools such as histograms and other charts to effectively interpret the collected data. As depicted in Figure~\ref{fig3_online}, two methods have been provided for real-time data monitoring:

\begin{figure}[h]
    \centering
    \includegraphics[width=0.8\linewidth]{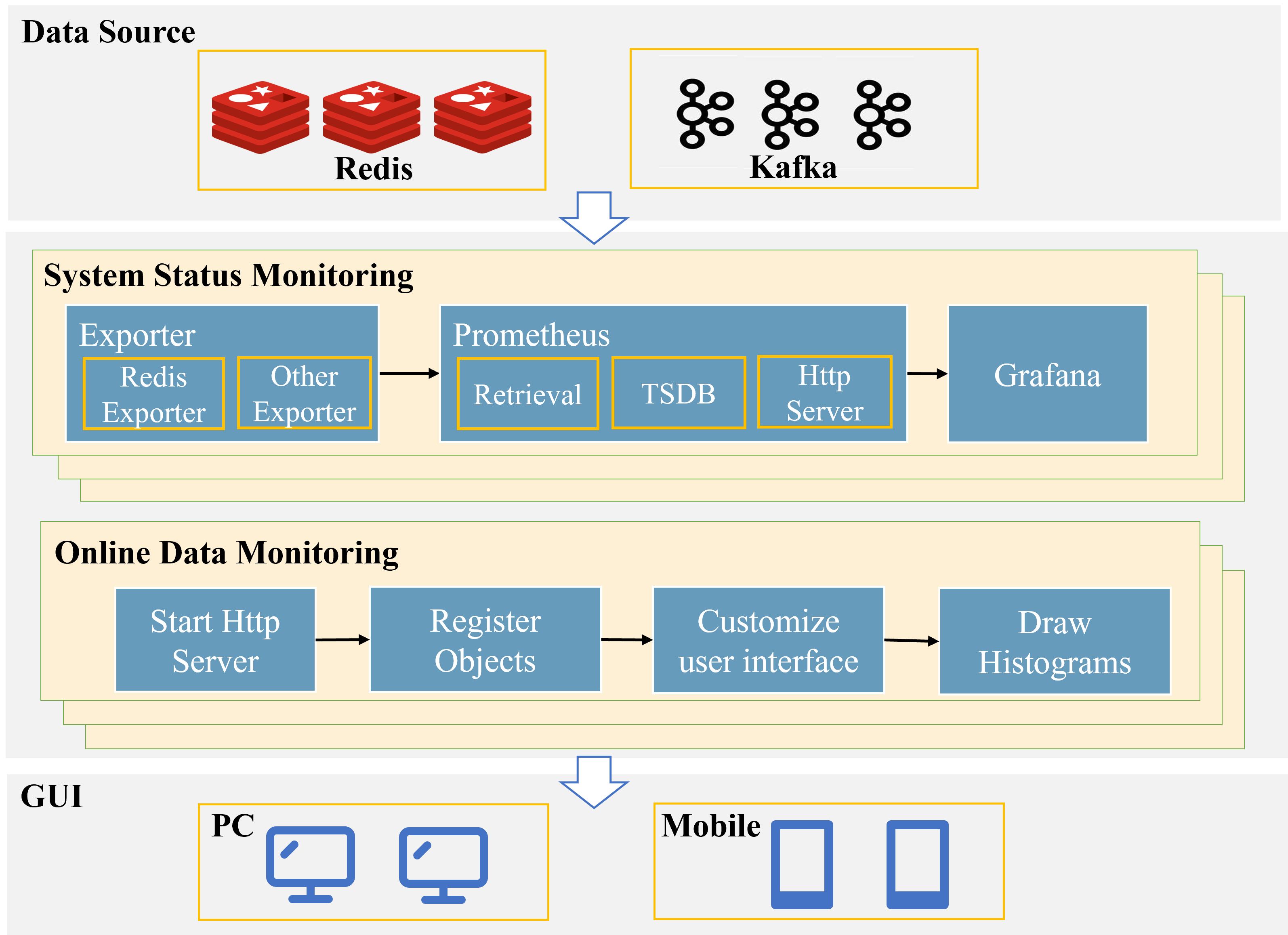}
    \caption{The architecture of TAO DAQ online monitoring system}
    \label{fig3_online}
\end{figure}

-	System status monitoring

The system status monitoring is implemented through the Exporter, Prometheus \cite{Prometheus} and Grafana \cite{grafana} approach. This combined approach provides an effective solution for visualizing the running status of the system. Exporters serve as agents that collect and export metrics from various systems and services, making the data available for Prometheus to scrape. Prometheus, known for its dimensional data model and flexible query language, stores the data from the exporters in a time series database (TSDB) and offers robust monitoring and alerting capabilities. Grafana, a popular open-source platform for data visualization, enables users to create visually appealing dashboards and charts from the data stored in Prometheus, providing a comprehensive solution for monitoring, collecting, storing, querying, and visualizing metrics and time-series data.

-	Online data monitoring

An online real-time data visualization system based on ROOT \cite{jsroot} has been designed and developed  \cite{zhangsh}. ROOT is an open-source data analysis framework widely used in high-energy physics, providing a comprehensive set of tools for data storage, manipulation, and analysis, including histogram, fitting, and visualization. Its flexibility and extensive capabilities make it a popular choice for handling large-scale data analysis tasks in research and academic settings. Therefore, to accommodate the preferences of physicists, a ROOT-based monitoring method has been provided. Furthermore, to offer a user-friendly graphical interface, THttpServer has been implemented to provide remote Hypertext Transfer Protocol (HTTP) access and enable an HTML/JavaScript user interface. This approach allows users to request and visualize data in a web browser without generating root files. Users can obtain the corresponding data from the message service interface of the online software and upload it to the visualization system to display the respective histograms.

At present, the online software has been successfully integrated with the data flow system and is undergoing functional testing and validation.

\subsection{Integration test with each subsystem}\label{subsec2}

As illustrated in Table \ref{tab2_testPlat}, the test bench for each subsystem has been constructed in the Institute of High Energy Physics (IHEP). All three detectors have undergone multiple tests to validate the interfaces and basic functions, including register configuration, data readout, data format verification, data processing and data storage. The data flow system has also completed the system implementation, providing corresponding data readout interfaces and software trigger algorithms. Additionally, the online software component has implemented a WEB Graphical User Interface (GUI) for real-time online monitoring and run control.

\begin{table}[h]
\caption{The full chain test bench in IHEP}\label{tab2_testPlat}%
\begin{tabular}{@{}llll@{}}
\toprule
Detector & Readout chain  & Number of channels connected & Data Type\\
\midrule
CD (1:1 prototype)    & SiPM+FEB+FEC+CU+DAQ  & $\sim$ 60 channels  & T/Q hits  \\
WT    & PMT+ABC+GCU+DAQ   & $\sim$ 300 channels  & T/Q hits  \\
TVT    & SiPM+FEB+FEC+TGU+DAQ   & 4 channels  & T/Q hits  \\
\botrule
\end{tabular}
\end{table}

For CD, we are currently utilizing a 1:1 prototype for systematic testing and validation. Final components are used together with pre-production components to test key installation procedures with a part of the condition limitations in Taishan. At the same time, the performance of each subsystem can be tested in the 1:1 prototype, avoiding big issues and saving time on site. The full-chain testing system, including the SiPM, FEB, FEC, CU, and DAQ, has been deployed. The DAQ system is connected to approximately 60 channels, with T/Q data taking. The data flow system has been enhanced with an online data display, which has successfully taken approximately 2 GB of data for analysis. The DAQ system has been running consistently, providing crucial assurance for electronics testing and data analysis in the 1:1 prototype.


For TVT, a complete chain test system has also been built, consisting of PS, FEB, FEC, TGU, and DAQ. The DAQ system is currently connected to 4 FEC channels for T/Q data acquisition. The data flow system with layer coincidence trigger algorithm has successfully taken data from an FEC board and can run stably for two hours.

\begin{figure}[h]
    \centering
    \includegraphics[width=\textwidth]{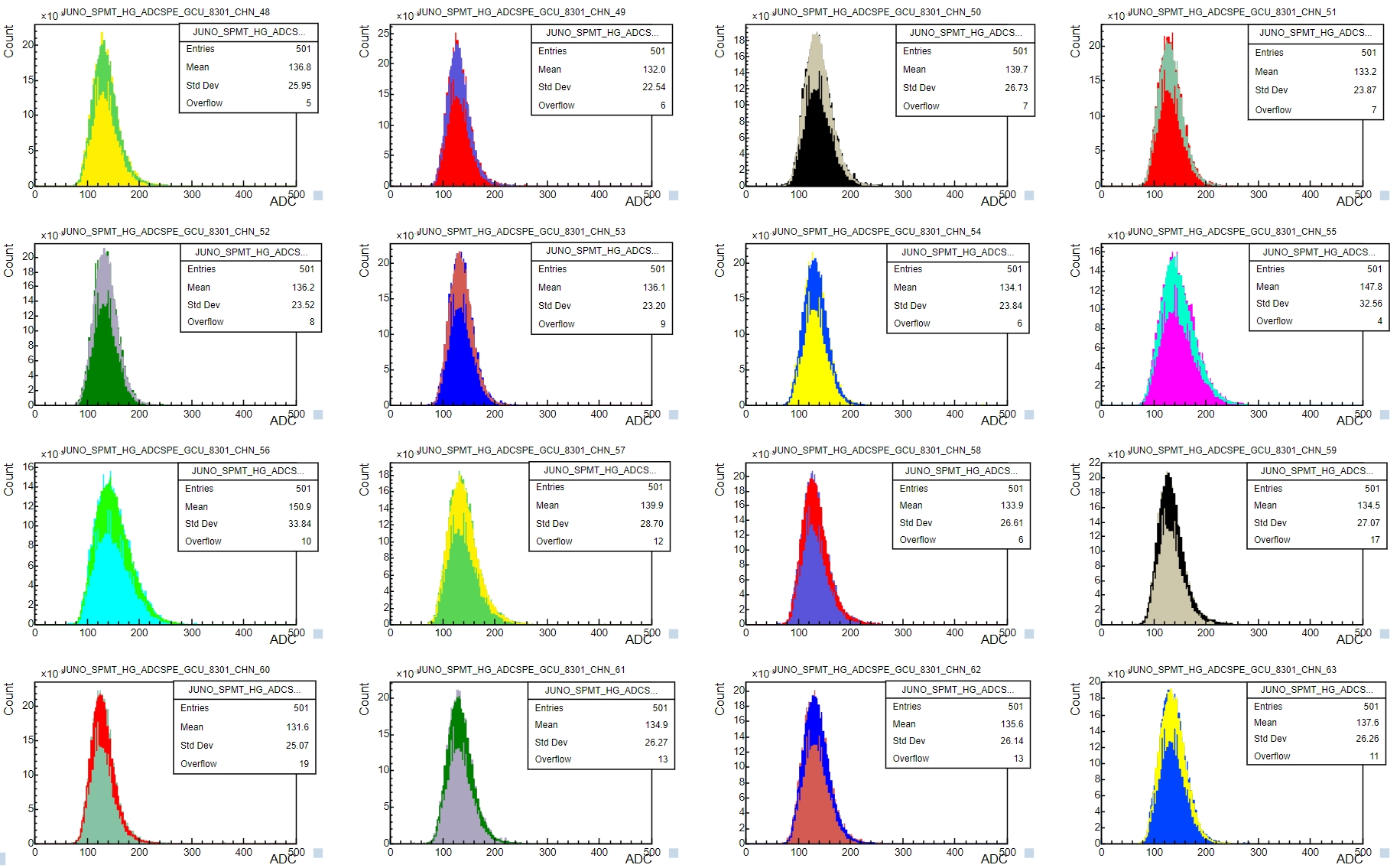}
    \caption{WT charge spectra online display during testing}
    \label{fig5_WT}
\end{figure}

For WT, the DAQ prototype has been deployed on several test platforms and can run stably to help with system diagnostics. During the test period, approximately 300 channels of data were acquired simultaneously, and a real-time energy spectrum display was incorporated into the system. Figure~\ref{fig5_WT} shows the real-time charge spectra of a GCU with 16 channels during an integration test. The system has undergone several comprehensive full-chain integration tests, collecting approximately 5 terabytes of data for analysis.

In addition, owing to the display needs of different detectors, a ROOT-based general online data visualization system has been deployed to provide a convenient integration solution. Users only need to modify the corresponding display parameters in the configuration file to complete online real-time histogram monitoring for different detectors.

\section{Performance test}\label{sec5}

The data acquisition system depends on the front-end electronics system to provide raw data for the system's normal operation. However, as the electronics system is currently in the debugging phase, a dummy data source software has been developed for each detector system to validate the performance of the data acquisition system at full-scale data volume.

\begin{figure}[h]
    \centering
    \includegraphics[width=0.8\linewidth]{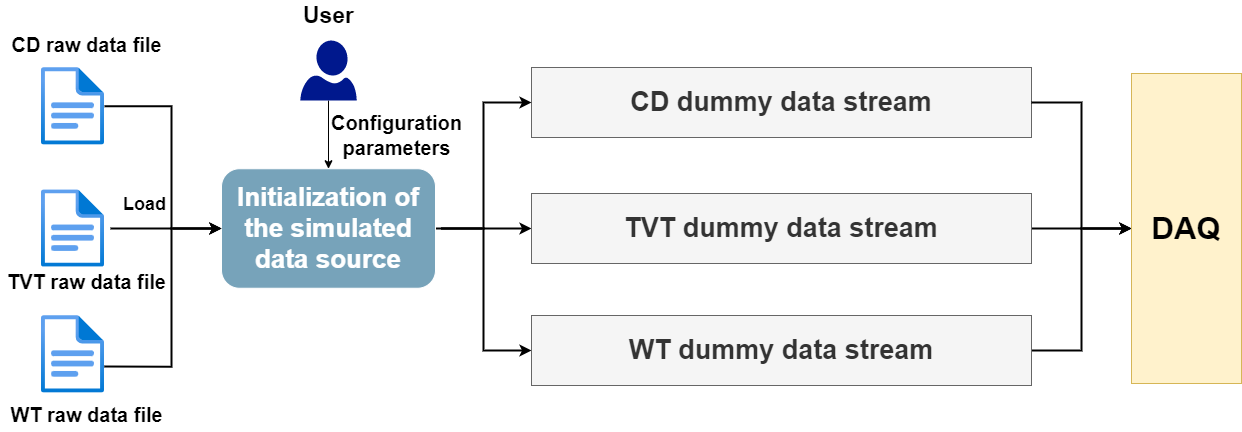}
    \caption{The implementation process of the dummy data source software}
    \label{dummysource}
\end{figure}

The front-end electronics of the TAO experiment mainly generate three types of data streams to transmit to the data acquisition system: CD T/Q, TVT T/Q, and WT T/Q data. To simulate real-world applications, multi-threading technology is used to allocate a data transmission thread for each electronic or trigger board. The specific implementation process is shown in Figure~\ref{dummysource}. Initially, the software reads the raw data files of each detector, generates the corresponding data format based on the user configuration parameters, and stores it in memory. Following initialization, the software simulates data transmission from the electronic boards to the DAQ system at a configurable data rate via TCP. During performance testing, all processes in the data flow system are deployed on a single server, with the server specifications detailed in Table~\ref{tab2_envir}.

\begin{table}[h]
\caption{Test Environment Main Parameters}\label{tab2_envir}%
\begin{tabular}{@{}llll@{}}
\toprule
OS & CentOS 7.9 \\
\midrule
Kernel version  & 3.10.0-1160.el7.x86\_64   \\
Compiler    & GCC8.3   \\
CPU Model   & Intel(R) Xeon(R) Gold 6226R CPU @ 2.90GHz (16cores x 2)   \\
CPU Cores   & 64   \\
Hyperthreading   & on   \\
\botrule
\end{tabular}
\end{table}

\subsection{Data readout}\label{subsec2}


To validate the data readout performance and stability of the data flow system for TAO experiments, the dummy data source is employed to send raw data from three detectors to the data flow system. After that, we perform a dummy mixed data test, including CD T/Q, TVT T/Q, and WT T/Q data. The specific sending frequency for each detector is consistent with that specified in the section \ref{sec2}. 

\begin{figure}[h]
    \centering
    \includegraphics[width=1\linewidth]{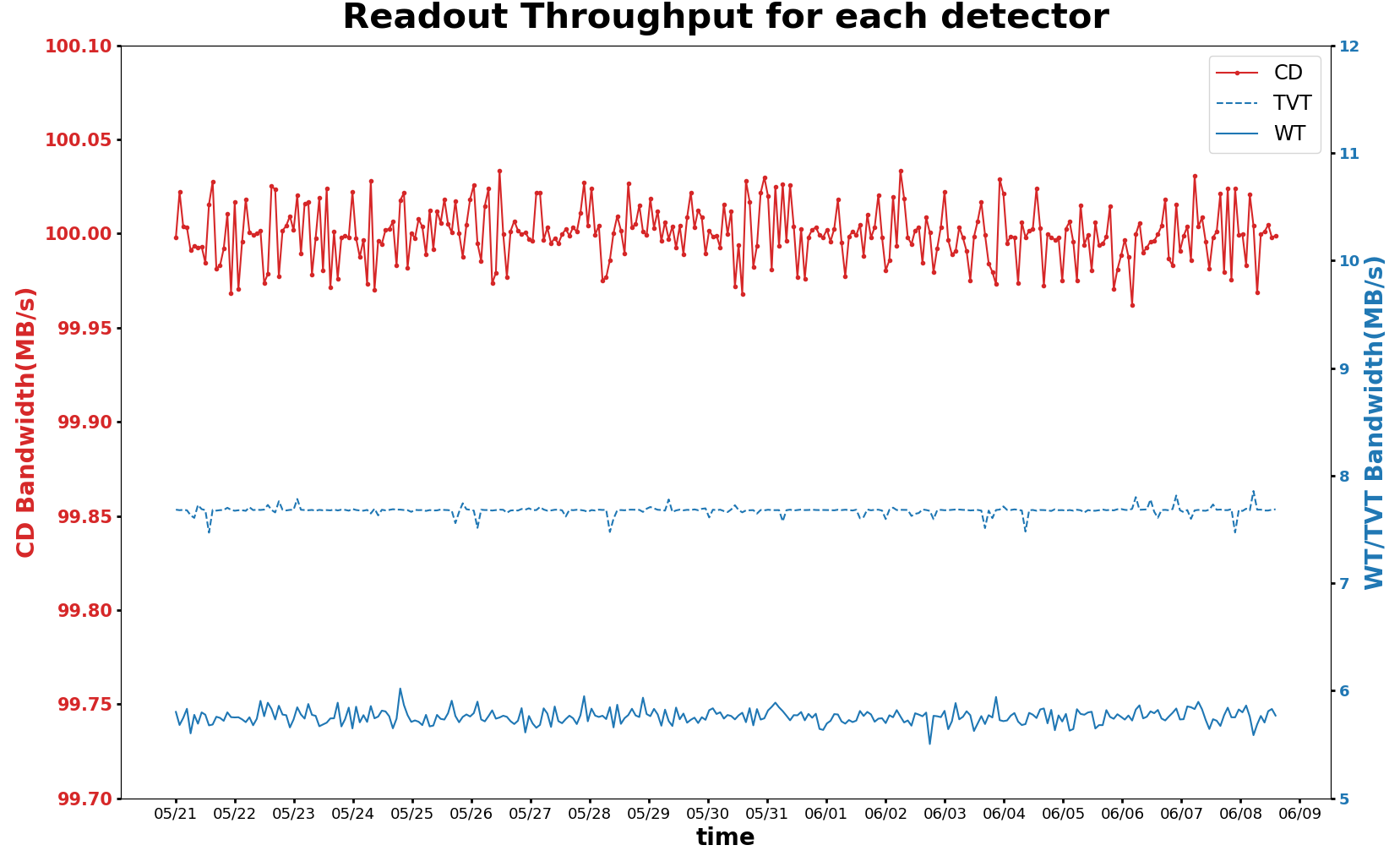}
    \caption{Readout bandwidth with dummy full-scale data volume}
    \label{fig6_readoutBw}
\end{figure}

\figureautorefname~\ref{fig6_readoutBw} demonstrates the readout bandwidths of the full-chain data flow system over eighteen days, and the software was interrupted due to a power outage. For the red curve labeled on the CD, the sending frequency is 1 kHz, with a total data bandwidth of 800 Mbps. For the blue curves, representing the WT and TVT, the total readout bandwidth is approximately 140 Mbps. The total bandwidth is sufficient to satisfy the readout requirements of the three detectors. It also can be observed that, when simulating the full-scale data volume, the data flow system can run stably. In addition, to increase the redundancy of the data flow system, the data rate of the three detectors is increased to 1.5 times the design specification, such as the data rate of the CD being set to 1.5 kHz for the full chain data flow system test. The data readout bandwidth for the three detectors matches the settings of the dummy data source, ensuring redundancy in the system.

\subsection{Data processing}\label{subsec2}
The data processing for the TAO experiment involves two key components: software trigger and data compression, each requiring a separate performance test. For the software trigger, the algorithm performance for WT and TVT is detailed in Table \ref{tab3_software}. Five key parameters affect the results of the software trigger, including channel numbers, hit rate, data window, trigger window, and threshold. Therefore, based on the specific parameters provided by the TAO VETO system, the execution time and CPU core usage of the algorithm were obtained with the input of a random distribution of T/Q data, thus reflecting the performance of the software trigger algorithm. In terms of data compression, we researched popular lossless compression algorithms available, considering factors such as compression time and ratio. The compression performance of different algorithms was evaluated using the same 2 Gigabyte dataset from integration testing, and the results are summarized in Table \ref{tab4_compression}. The test results demonstrate the feasibility of the processing algorithms, as the algorithms consume relatively few CPU resources. This means that in practical applications, the algorithms can efficiently utilize computing resources.

\begin{table}[h]
\caption{Performance of software trigger algorithm for TAO}\label{tab3_software}
\begin{tabular*}{\textwidth}{@{\extracolsep\fill}lccccccc}
\toprule%
& \multicolumn{5}{@{}c@{}}{Algorithm Parameters} & \multicolumn{2}{@{}c@{}}{Trigger Performance} \\\cmidrule{2-6}\cmidrule{7-8}%
Detector & \makecell{Channel\\Numbers} & \makecell{Hit\\Rate} & \makecell{Data\\Window} & \makecell{Trigger\\Window} & \makecell{Threshold} & \makecell{Algorithm\\Time} & \makecell{CPU\\Cores} \\
\midrule
WT  & 300 & 5.1 kHz & 500 ns & 300 ns & 5 & 1257 $\mu$s & 1$(\sim 0.2$)\\
TVT  & 4 layers & 80 kHz  & 300 ns & 100 ns & 2 & 981 $\mu$s & 1$(\sim 0.1$)\\
\botrule
\end{tabular*}
\end{table}

\begin{table}[h]
\caption{Performance of data compression}\label{tab4_compression}%
\begin{tabular}{@{}llll@{}}
\toprule
Algorithm & File Size After compression &   Algorithm Time(s) & Compression Ratio \\
\midrule
bzip2  & 647 MB & 122.679267 & 3.16   \\
gzip    & 833 MB & 111.815730 & 2.45   \\
zip   & 833 MB & 106.591130 & 2.45   \\
LZ4 HC -9   & 1187 MB & 80.061533 & 1.88   \\
compression   & 1046 MB & 32.536046 & 1.95   \\
lz4   & 1432 MB & 21.810871 & 1.43   \\
\botrule
\end{tabular}
\end{table}

\section{Hardware architecture design}\label{sec5}

\begin{figure}[h]
    \centering
    \includegraphics[width=0.8\linewidth]{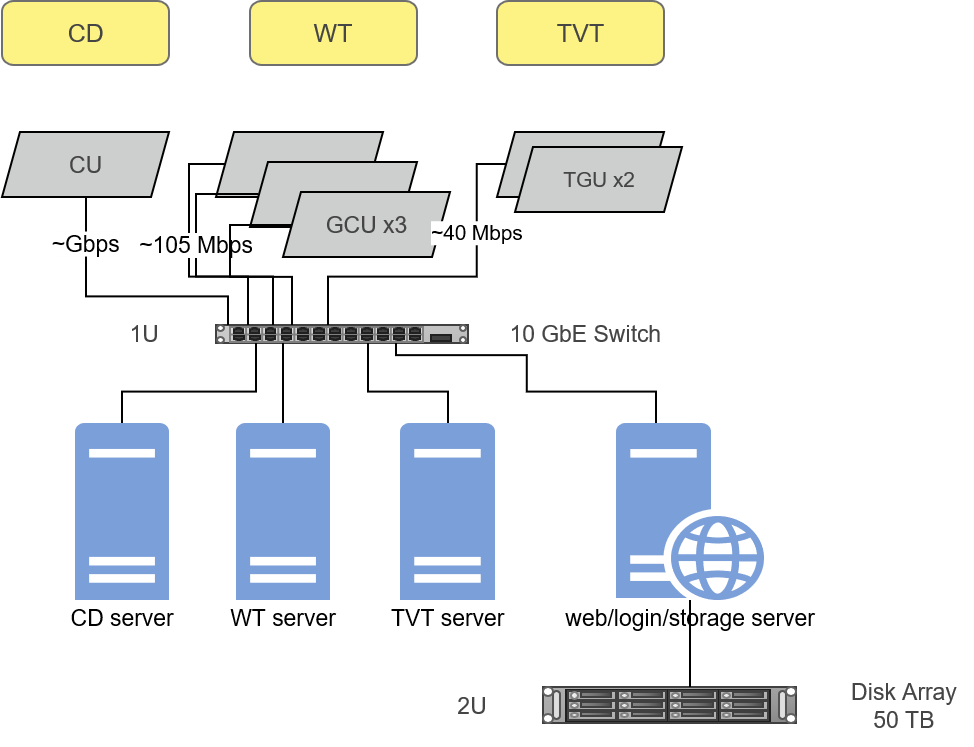}
    \caption{The hardware architecture of the TAO DAQ system}
    \label{hardarchitecture}
\end{figure}

Considering the modular design and high integration of the TAO DAQ system, a general and readily scalable hardware deployment scheme has been designed. The hardware architecture, as shown in Figure~\ref{hardarchitecture}, will be based on three key components: commercial computing servers, network devices, and storage devices. According to the data flow system architecture, the entire server cluster is divided into four parts that are responsible for readout, assembly, processing, and storage. Furthermore, other functions and services such as web and login will be utilized in the idle server. This approach enables a more streamlined and effective data flow system, enhancing the cluster's capabilities to handle various tasks and services.

The data transfer from the electronics or trigger system to the data acquisition system will be handled using standard Ethernet. This will be managed by a multi-port 10 Gigabit Ethernet switch to facilitate server communications. 

Additionally, the data generated by the TAO experiment will be temporarily stored in a 2U disk array with a capacity of approximately 50 TB, which can cache data for six weeks. 

\section{Conclusion}\label{sec7}
The core functionality of the TAO DAQ system has been designed and developed based on the established JUNO DAQ framework. Through successful application in laboratory tests, the system's structural design has been validated, affirming the interface usability of the data flow system. 

The flexible and scalable design of the DAQ system plays a significant role in the TAO experiment. As the experiment is currently undergoing pre-installation testing, comprehensive commissioning is ongoing. The system will be continuously refined and optimized based on the test results, ensuring its readiness for the experimental phase.

\bmhead{Acknowledgements}

We would like to extend our sincere thanks to the members of the JUNO and TAO collaboration groups for their invaluable support and assistance. They provided hardware equipment and assisted in completing the integration for the smooth progress of this project. Beyond that, this work is supported by the Strategic Priority Research Program of the Chinese Academy of Sciences, Grant No. XDA10000000 and XDA10010700, State Key Laboratory of Particle Detection and Electronics (SKLPDE) Grant No. SKLPDE-ZZ-202305 and National Key R\&D Program of China by the Ministry of Science and Technology (MoST), Grant No. 2022YFA1602002. On behalf of all authors, the corresponding authors state that there is no conflict of interest.


\bibliography{sn-bibliography}

\end{document}